\documentclass[preprint,12pt,sort&compress]{elsarticle}
\usepackage{graphicx}

\begin{document}
\begin{frontmatter}

\title{Appearing (disappearing) lumps and rogue lumps of the two-dimensional  vector
Yajima--Oikawa system}

\author[lab1,lab2]{N.V.~Ustinov\corref{cor1}}
\address[lab1]{Department of Informational Technologies, Kaliningrad Institute of Management, 236001 Kaliningrad, Russia}
\address[lab2]{Department of Photonics and Microwave Physics, Lomonosov Moscow State University, 119991 Moscow, Russia}
\ead{n\_ustinov@mail.ru}
\cortext[cor1]{Corresponding author}

\begin{abstract}
The solutions of the two-dimensional multicomponent Yajima--Oikawa system that have the functional arbitrariness are constructed by using the Darboux transformation technique. 
For the zero and constant backgrounds, different types of solutions of this system, including the lumps, line rogue waves, semi-rational solutions and their higher-order counterparts, are considered. 
Also, the generalization of the lump solutions (namely, appearing or disappearing lumps) is obtained in the two-component case under the special choice of the arbitrary functions. 
Then, the suitable ansatz is used to find the further generalization of these lumps (appearing-disappearing lumps or rogue lumps). 
\end{abstract}


\end{frontmatter}

\section{Introduction}

The considerable attention was paid in the recent years to an investigation of the 
two-dimensional multicomponent Yajima-Oikawa (YO) system 
\cite{OMO,RKLG,KVSL,SK,KVL,KKT,CCFM_1,CCFM_2,CCFMa,RPHK,SU4}.
This system comprises multiple (say $N$) short-wave components and a single long-wave one. 
It generalizes the scalar ($N=1$) two-dimensional YO system \cite{ZMNP} and is often called the 2D coupled long-wave--short-wave resonance interaction system.  

The two-dimensional multicomponent YO systems can arise in different physical contexts. 
In particular, the two-component ($N=2$) system and the multicomponent one were derived by applying the reductive perturbation method in Refs.~\cite{OMO} and \cite{KVL}, respectively, as the governing equation for the interaction of dispersive waves in a weak Kerr-type nonlinear medium in the small amplitude limit. 
In these systems, the short waves propagate in anomalous dispersion regime while the long wave propagates in the normal dispersion regime. 
Also, a generation of the terahertz radiation by optical pulses in a medium of asymmetric quantum particles is described under the quasi-resonance conditions by the two-dimensional two-component YO system \cite{SU4}. 
It is worth to note that the two-dimensional scalar YO system was derived for a two-layer fluid model by using the multiple scale perturbation method \cite{OOF}. 
Interestingly, this system can be also deduced from the governing equations for 
two-dimensional two-wave interaction \cite{OOS,SKEMS} by means of the reductive perturbation method (see Refs.~\cite{OMO,TWKBFB,MWT,CL}). 

It is well-known \cite{APT} that the solutions of the multicomponent systems exhibit some novel properties that have not been observed in the scalar (single-component) counterpart. 
It was found, in particular, that the bright solitons of the two-dimensional two-component YO system undergo the energy-sharing (inelastic) collisions \cite{KVSL}. 
The unusual properties of the interaction of the multimode dromions of this system was revealed in Ref.~\cite{RKLG}. 
Note that the solutions of the one-dimensional multicomponent YO system display also some specific features \cite{KST,CCKGD,CCFM}. 

Recently, the study of the rational solutions of the two-dimensional equations and their generalizations (such as the lump-type solutions and the semi-rational ones) has attracted a lot of attention \cite{CCFM_2,RPHK,EPV,KRL,DM_1,OY_1,OY_2,DM_2,WD,CZ,BWK,EDDACD,CGMB,W,
TD,MZD,WZ,WDL,RPH,AERS,LWCL,L,CMH,MS,HRA,PTZ,MZ,RMCH,ZWZF,WTQDZZ,LWW,TDY,CW}. 
Some of these solutions are suitable in describing the behavior of rogue (or freak) waves that were intensively investigated in the last decades 
\cite{OOS,KPS,ORBMA,DDEG,Mih,CBSGM,CCMFK}. 
In particular, the rogue wave solutions were found for the Kadomtsev--Petviashvili equation 
\cite{DM_1,DM_2,MS}, Davey--Stewartson eqiation \cite{OY_1,OY_2},  
Nizhnik--Novikov--Veselov equation \cite{AERS,WD} and some other (2+1)-dimensional  equations. 
As a rule, the solutions in the form of rogue waves are obtained from the lump solutions by a reduction of variables. 

The rational and semi-rational solutions of the two-dimensional multicomponent YO system were investigated in Refs.~\cite{CCFM_2} and \cite{RPHK}, respectively. 
The rational solutions were obtained by means of the bilinear method and include the fundamental (simplest) and general (multi- and higher-order) lumps and line rogue waves.
It was shown that the fundamental lumps and rogue waves have three different patterns: bright, intermediate and dark states \cite{CCFM_2}. 
The fundamental semi-rational solutions obtained also by the bilinear method can describe the fission of a dark soliton into a lump and a dark soliton or the fusion of one lump and one dark soliton into a dark soliton \cite{RPHK}. 
The nonfundamental semi-rational solutions were shown to fall into three subclasses: higher-order, multi- and mixed-type semi-rational solutions.

The rich structure of the solutions of the two-dimensional multicomponent YO system is due to its higher-dimensional and multi-component nature. 
Being motivated by this, we exploit the Darboux transformation (DT) technique 
\cite{MaSa,GHZ} to obtain the solutions of this system. 
Note that the DT technique was applied to the multicomponent YO systems in the 
one-dimensional case in Refs.~\cite{SU2,SU3,C,CGSC}. 
 
The paper is organized as follows. 
The overdetermined system of linear equations of the two-dimensional multicomponent YO system of the general form is given in Section~2. 
The DT technique is applied also in this section to the systems considered. 
This technique allows us to construct the solutions of the two-dimensional multicomponent YO system that contain the arbitrary functions. 
In Section~3, the solutions of this system on the zero and constant backgrounds are discussed. 
The case of the two-component two-dimensional YO system is studied in details in Section~4. The solutions in the form of the appearing (disappearing) lumps are obtained in this section by taking the arbitrary functions in a special manner. 
Then, the ansatz is used to construct the generalization of these solutions in the form of appearing-disappearing lump (or rogue lump). 
The estimation of the lifetime of this lump is given. 
The main results are summarized in Section~5.

\section{Overdetermined linear system and Darboux transformation}

Consider the two-dimensional multicomponent YO system
\begin{equation}
\begin{array}{c}
{\displaystyle i\left(\frac{\partial\varphi_n}{\partial y}+
\frac{\partial\varphi_n}{\partial x}\right)=-\frac{\partial^2\varphi_n}
{\partial\tau^2}-u\varphi_n,}_{\mathstrut}\\
{n=1,\dots,N,}_{\mathstrut}^{\mathstrut}\\
{\displaystyle\frac{\partial u}{\partial y}=\frac{\partial}{\partial\tau}
\sum\limits_{n=1}^N\sigma_n|\varphi_n|^2}^{\mathstrut}.
\end{array}
\label{N_YO}
\end{equation}
Here $\varphi_n$ is the $n$th short-wave component, $u$ is the long-wave one, $\sigma_n=\pm1$ ($n=1,\dots,N$).

If the wave components are independent of variable $x$, then Eqs.~(\ref{N_YO}) are reduced to the one-dimensional multicomponent YO system. 
This system has various physical applications \cite{SU2,SU3,M,MPK,SU1,BS}. 
In particular, it describes in the case $N=2$ and $\sigma_1=\sigma_2=1$ the propagation of the vector electromagnetic and acoustic pulses \cite{SU2,SU3,SU1,BS}. 
In these physical applications, the variable $\tau$ plays the role of the dimensionless "local" time, while the variable $y$ is the dimensionless spatial coordinate. 
Here, such an interpretation of the independent variables is understood also. 
Besides, this corresponds to the roles of the independent variables in the two-dimensional 
two-component YO system studied in Ref.~\cite{SU4} under a consideration of the terahertz radiation generation by the optical pulses in a medium of asymmetric quantum particles.

Eqs.~(\ref{N_YO}) are integrable by the inverse scattering transformation method \cite{M} and admits a representation as the compatibility condition of the overdetermined system of linear equations 
\begin{equation}
\begin{array}{c}
{\displaystyle\frac{\partial^2\psi_1}{\partial\tau^2}=-i\left(
\frac{\partial\psi_1}{\partial y}+\frac{\partial\psi_1}{\partial x}\right)-
u\psi_1,}_{\mathstrut}\\
{\displaystyle\frac{\partial\psi_{n+1}}{\partial\tau}=
\frac{\sigma_{n+1}}{2}\varphi_n^*\psi_1,}^{\mathstrut}_{\mathstrut}\\
{n=1,\dots,N,}^{\mathstrut}
\end{array}
\label{le1}
\end{equation}
and 
\begin{equation}
\begin{array}{c}
{\displaystyle\frac{\partial\psi_1}{\partial y}=-\sum\limits_{n=1}^N\varphi_n\psi_{n+1}
,}_{\mathstrut}\\
{\displaystyle\frac{\partial\psi_{n+1}}{\partial y}+\frac{\partial\psi_{n+1}}{\partial x}=
\frac{i\sigma_n}{2}\left(\varphi_n^*\frac{\partial\psi_1}{\partial\tau}-
\frac{\partial\varphi_n^*}{\partial\tau}\psi_1\right),}^{\mathstrut}_{\mathstrut}\\
{n=1,\dots,N,}^{\mathstrut}
\end{array}
\label{le2}
\end{equation}
where $\psi_k=\psi_k(\tau,y,x)$ ($k=1,\dots,N+1$) is the solution of Eqs.~(\ref{le1}) and (\ref{le2}).

Let $\chi_k=\chi_k(\tau,y,x)$ ($k=1,\dots,N+1$) be a solution of the overdetermined system (\ref{le1}) and (\ref{le2}). 
Then the differential 1-form 
$$
d\,\delta(\chi,\psi)=\delta_{\tau}(\chi,\psi)d\tau+\delta_y(\chi,\psi)dy+
\delta_x(\chi,\psi)dx,
$$
where 
$$
\delta_{\tau}(\chi,\psi)=\chi_1^*\psi_1, 
$$
$$
\delta_y(\chi,\psi)=-2\sum\limits_{n=1}^N\sigma_n\chi_{n+1}^*\psi_{n+1}, 
$$
$$
\delta_x(\chi,\psi)=i\left(\chi_1^*\frac{\partial\psi_1}{\partial\tau}-
\frac{\partial\chi_1^*}{\partial\tau}\psi_1\right)+2\sum\limits_{n=1}^N\sigma_n\chi_{n+1}^*\psi_{n+1}, 
$$
is closed; i.e., for a contour $\Gamma$ connecting the points $(\tau_0,y_0,x_0)$ and $(\tau,y,x)$, an integral 
\begin{equation}
\delta(\chi,\psi)=\int\limits_{\Gamma}d\,\delta(\chi,\psi)+C
\label{delta}
\end{equation}
($C$ is a constant) depends only on the initial and final points and does not depend on a specific choice of the contour. 

Let us apply the DT technique to obtain the solutions of the two-dimen\-sional multicomponent YO system (\ref{N_YO}). 
The overdetermined system (\ref{le1}) and (\ref{le2}) of linear equations is covariant with respect to the DT $\psi_k\to\psi_k[1]$ ($k=1,\dots,N+1$), $\varphi_n\to\varphi_n[1]$ ($n=1,\dots,N$), $u\to u[1]$, where the transformed quantities are defined in the following manner \cite{SU4}:
\begin{equation}
\psi_k[1]=\psi_k-\frac{\delta(\chi,\psi)}{\delta(\chi,\chi)}\chi_k,\ k=1,\dots,N+1, 
\label{DT_psi}
\end{equation}
\begin{equation}
\varphi_n[1]=\varphi_n-2\frac{\chi_{n+1}^*\chi_1}{\delta(\chi,\chi)},\ n=1,\,\dots,\,N,
\label{DT_phi}
\end{equation}
\begin{equation}
u[1]=u+2\frac{\partial^2}{\partial\tau^2}\log\delta(\chi,\chi). 
\label{DT_u}
\end{equation}
Relations (\ref{DT_phi}) and (\ref{DT_u}) define a new ("dressed") solution of the system (\ref{N_YO}), while the expressions (\ref{DT_psi}) give corresponding solutions of the overdetermined system of linear equations. 

Having expressions (\ref{DT_psi}) for the solutions of the transformed overdetermined system, we can perform the iterations of the DT considered. 
Let $\chi^{(l)}_k=\chi^{(l)}_k(\tau,y,x)$ ($k=1,\dots,N+1$, $l=2,\dots,L$) be the solutions of the overdetermined system (\ref{le1}) and (\ref{le2}). 
Taking into account the identity
$$
\delta(\chi^{(l)}[1],\psi[1])\delta(\chi,\chi)=
\left|
\begin{array}{ll}
\delta(\chi^{(l)},\psi)&{\delta(\chi,\psi)}_{\mathstrut}\\
\delta(\chi^{(l)},\chi)&{\delta(\chi,\chi)}_{\mathstrut}
\end{array}
\right|,
$$
we obtain after the $L$-fold iteration of the DT (\ref{DT_psi})--(\ref{DT_u}) the following expressions for the transformed solutions of the two-dimensional multicomponent YO system (\ref{N_YO}):
\begin{equation}
\varphi_n[L]=\varphi_n-\frac{2}{D[L]}\sum\limits_{l,m=1}^{L}D[L]^{(l,m)}
{\chi_{n+1}^{(l)}}^*\chi_1^{(m)},\ n=1,\,\dots,\,N,
\label{DT_phi_L}
\end{equation}
\begin{equation}
u[L]=u+2\frac{\partial^2}{\partial \tau^2}\log D[L],
\label{DT_u_L}
\end{equation}
where 
$$
D[M]=\left|
\begin{array}{ccc}
\delta({\chi^{(1)}},\chi^{(1)})&\ldots&\delta({\chi^{(1)}},\chi^{(M)})\\
\vdots&\ddots&\vdots\\
\delta({\chi^{(M)}},\chi^{(1)})&\ldots&\delta({\chi^{(M)}},\chi^{(M)})
\end{array}
\right|,
$$
$D[L]^{(l,m)}$ is the algebraic complement of the element in the $l$th row and $m$th column of the matrix of the determinant $D[L]$. 
Here we put $\chi_k=\chi^{(1)}_k$ ($k=1,\dots,N+1$). 

\section{The zero and constant backgrounds}

The DT formulas (\ref{DT_phi}), (\ref{DT_u}) and (\ref{DT_phi_L}), (\ref{DT_u_L}) allows us to construct solutions of the two-dimensional YO system (\ref{N_YO}), which have the functional arbitrariness. 
To illustrate this we consider the zero and constant backgrounds. 

Let the initial solution of the YO system (\ref{N_YO}) be the zero background:
$$
\varphi_1=\dots=\varphi_N=u=0.
$$
A solution of the overdetermined system (\ref{le1}) and (\ref{le2}) can be chosen in this case in the form 
\begin{equation}
\chi_1=\int\!\!\!\!\!\int\limits_{D\,\ \ }\!\!\sum\limits_{j=0}^{\infty}\Lambda_j(\mu)
\frac{\partial^j}{\partial\mu^j}\,{\rm e}^{\Delta_0(\mu)}d\mu_Rd\mu_I,
\label{chi_0_1}
\end{equation}
\begin{equation}
\chi_{n+1}=f_n(y-x),\ n=1,\dots,N,
\label{chi_0_j}
\end{equation}
where $\mu$ is a complex parameter, $\mu_R=\Re(\mu)$, $\mu_I=\Im(\mu)$, $D$ is a domain on the parameter $\mu$ plane, 
$$
\Delta_0(\mu)=\mu\tau+i\mu^2x+f_0(\mu),
$$
$\Lambda_j(\mu)$ ($j=0,1,2,\dots$), $f_0(\mu)$ and $f_n(y-x)$ ($n=1,\dots,N$) are arbitrary functions of their arguments. 

In the simplest case, when 
$$
\Lambda_0(\mu)=\delta(\mu_R-\lambda)\delta(\mu_I-\nu),\ \Lambda_j(\mu)=0,\ j\in\bf N,
$$
where $\delta(\mu_R-\lambda)$ and $\delta(\mu_I-\nu)$ are Dirac's delta functions, $\lambda$ and $\nu$ are real constants, such that $\lambda+i\nu\in D$, expression 
(\ref{chi_0_1}) is reduced to 
\begin{equation}
\chi_1={\rm e}^{\Delta_0(\lambda+i\nu)}.
\label{chi_0_1_s}
\end{equation}
Then, we have from Eq.~(\ref{delta})
\begin{equation}
\begin{array}{c}
{\displaystyle\delta(\chi,\chi)=\frac{1}{2\lambda}{\rm e}^{\Delta_0(\lambda+i\nu)+\Delta_0(\lambda+i\nu)^*}-\mbox{}}_{\mathstrut}\\
{\displaystyle2\int\limits_{y_0-x_0}^{y-x}\sum\limits_{n=1}^{N}\sigma_n|f_n(\zeta)|^2d\zeta+a_0,}^{\mathstrut}
\end{array}
\label{delta_0_s}
\end{equation}
where $a_0$ is a real constant. 

Substitution of the expressions (\ref{chi_0_j})--(\ref{delta_0_s}) into the DT formulas (\ref{DT_phi}) and (\ref{DT_u}) gives the simplest solution of the two-dimensional YO system (\ref{N_YO}) that has the functional arbitrariness. 
If $f_n(y-x)\sim\exp[\varepsilon(y-x)]$ ($n=1,\dots,N$; $\varepsilon$ is a real constant), and constant $a_0$ is defined properly, then this solution is a line soliton. 
Condition $\varepsilon=2\lambda\nu$, corresponds to the one-dimensional case. 
If, for example, $f_n(y-x)\sim1/\cosh[\varepsilon(y-x)]$ ($j=n,\dots,N$), and constant $a_0$ is chosen in a proper manner, then the solution of system (\ref{N_YO}) has localized short-wave components propagating along the lines $y-x=\rm const$.

Let us consider the initial solution of the YO system (\ref{N_YO}) in the form of the constant background:
\begin{equation}
\begin{array}{c}
{\varphi_n=a_n,\ n=1,\dots,N,}_{\mathstrut}\\
{u=0.}^{\mathstrut}
\end{array}
\label{cb}
\end{equation}
Without loss of generality, we assume here that the constants $a_n$ ($n=1,\dots,N$) are real and $a_N\ne0$. 
Then, a solution of the overdetermined system (\ref{le1}) and (\ref{le2}) can be represented in the following manner: 
\begin{equation}
\begin{array}{c}
{\displaystyle\chi_1=\int\!\!\!\!\!\int\limits_{D\,\ \ }\!\!\sum\limits_{j=0}^{\infty}\Lambda_j(\mu)\frac{\partial^j}{\partial\mu^j}\,{\rm e}^{\Delta(\mu)}d\mu_Rd\mu_I,
}_{\mathstrut}\\
{\displaystyle\chi_{n+1}=\frac{\sigma_na_n}{2}\int\!\!\!\!\!\int\limits_{D\,\ \ }\!\!\sum\limits_{j=0}^{\infty}\Lambda_j(\mu)\frac{\partial^j}{\partial\mu^j}\left(
\frac{{\rm e}^{\Delta(\mu)}}{\mu}\right)d\mu_Rd\mu_I+
f_n(\tilde y-\tilde x),}^{\mathstrut}_{\mathstrut}\\
n=1,\dots,N,
\end{array}
\label{chi}
\end{equation}
where 
$$
\Delta(\mu)=\mu\tilde\tau+i\mu^2\tilde x-\frac{\tilde y-\tilde x}{2\mu}
\sum\limits_{n=1}^{N}\sigma_na_n^2,
$$
$$
\tilde\tau=\tau-\tilde\tau_0,\ \tilde x=x-\tilde x_0,\ \tilde y=y-\tilde y_0,
$$
$\tilde\tau_0$, $\tilde x_0$ and $\tilde y_0$ are real constants, $\Lambda_j(\mu)$  ($j=0,1,2,\dots$) and $f_n(\tilde y-\tilde x)$ ($n=1,\dots,N-1$) are arbitrary functions of their arguments,
$$
f_N(\tilde y-\tilde x)=-\frac{1}{a_N}\sum\limits_{n=1}^{N-1}a_nf_n(\tilde y-\tilde x).
$$ 

Different types of the solutions of the two-dimensional YO system (\ref{N_YO}) are obtained by the substitution of the expressions (\ref{chi}) into the DT formulas (\ref{DT_phi}), (\ref{DT_u}).  
In particular, the line soliton corresponds to the following choice: 
$$
\begin{array}{c}
{\displaystyle\Lambda_0=\delta(\mu_R-\lambda)
\delta(\mu_I-\nu),}_{\mathstrut}\\
{\displaystyle\Lambda_j(\mu)=0,\ j=1,2,\dots,}^{\mathstrut}
\end{array}
$$
\begin{equation}
f_n(\tilde y-\tilde x)=0,\ n=1,\dots,N.
\label{0}
\end{equation}
This line soliton is dark. 

The lump and line rogue wave solutions \cite{CCFM_2}, semi-rational ones \cite{RPHK} are obtained if 
\begin{equation}
\begin{array}{c}
{\displaystyle\Lambda_0(\mu)=0,\ \Lambda_1(\mu)=\delta(\mu_R-\lambda)
\delta(\mu_I-\nu),}_{\mathstrut}\\
{\displaystyle\Lambda_j(\mu)=0,\ j=2,3,\dots,}^{\mathstrut}
\end{array}
\label{lc}
\end{equation}
and conditions (\ref{0}) are imposed

In the cases of the lump and line rogue wave solutions, the constant $C$ in 
Eq.~(\ref{delta}) is chosen in a such manner depending on the initial point of contour $\Gamma$ that $\delta(\chi,\chi)$ is the product of the exponential function with the polynomial of the second degree with respect to the independent variables. 
These solutions were studied in details in Ref.~\cite{CCFM_2}.
In the case of the semi-rational solutions, $\delta(\chi,\chi)$ is a sum of the constant and the product of the exponential function with the polynomial of the second degree. 
These solutions describe a fission of the line dark soliton into the lump and the line dark soliton or a fusion of one lump and one line dark soliton into the line dark soliton \cite{RPHK}.

The higher-order lump and line rogue wave solutions, higher-order semi-rational solutions are obtained if only one arbitrary function $\Lambda_m(\mu)$ ($m\in\bf N$, $m\ge2$) in Eqs.~(\ref{chi}) differs from the zero: 
$$
\Lambda_m(\mu)=\delta(\mu_R-\lambda)\delta(\mu_I-\nu). 
$$
The solutions describing an interaction of the line solitons, usual and/or higher-order lumps, line rogue waves and semi-rational solutions are constructed by using the formulas (\ref{DT_phi_L}), (\ref{DT_u_L}) of the iterated DT. 

Note that the conditions (\ref{0}) on the functions $f_n(\tilde y-\tilde x)$ ($n=1,\dots,N$) have to take place in the cases of the line solitons, lumps, line rogue waves and the semi-rational solutions. 
In the next section, these conditions will not be imposed. 
This will give an opportunity to construct the generalizations of the lumps. 

\section{The two-dimensional two-component YO system}

Consider the YO system (\ref{N_YO}) in the simplest vector case: $N=2$, $\sigma_1=\sigma_2=1$.  
Let the initial solution of this system be the constant background (\ref{cb}).
Then, the solution localized on $(x,y)$-plane can be obtained by using the DT formulas 
(\ref{DT_phi}), (\ref{DT_u}) if conditions (\ref{lc}) are valid, $\nu=0$ and arbitrary function  $f_1(\tilde y-\tilde x)$ is defined in the following manner: 
$$
f_1(\tilde y-\tilde x)=\frac{a_2}{2}\,P(\tilde y-\tilde x)\exp
\left(\frac{a_1^2+a_2^2}{2\lambda}\,(\tilde x-\tilde y)+i\theta(\tilde y-\tilde x)\right),
$$
where $P(\tilde y-\tilde x)$ is arbitrary function of its argument, 
$\theta(\tilde y-\tilde x)$ is arbitrary real function. 
In the simplest case 
$$
P(\tilde y-\tilde x)=r, 
$$
where $r$ is a constant, this gives us the solution of the form 
\begin{equation}
\displaystyle\varphi_n=a_n-\frac{2\rho_n}{\rho_++|r|^2{\rm e}^{-2\lambda\tilde\tau}},\ n=1,2, 
\label{al_phi}
\end{equation}
\begin{equation}
u=2\frac{\partial^2}{\partial \tau^2}\log\left(\rho_++|r|^2{\rm e}^{-2\lambda\tilde\tau}\right),
\label{al_u}
\end{equation}
where 
$$
\rho_1=a_1(\rho_--2i\tilde x)+a_2r^*\rho_0{\rm e}^{\xi},
$$
$$
\rho_2=a_2(\rho_--2i\tilde x)-a_1r^*\rho_0{\rm e}^{\xi},
$$
$$
\rho_{\pm}=\left(\tilde\tau+\frac{a_1^2+a_2^2}{2\lambda^2}\,(\tilde y-\tilde x)\right)^2+4\lambda^2\tilde x^2\pm\frac{1}{4\lambda^2},
$$
$$
\rho_0=\tilde\tau+\frac{a_1^2+a_2^2}{2\lambda^2}\,(\tilde y-\tilde x)+2i\lambda\tilde x+
\frac{1}{2\lambda},
$$
$$
\xi=-\lambda\tilde\tau+i\lambda^2\tilde x-i\theta(\tilde y-\tilde x).
$$ 

Solution (\ref{al_phi}), (\ref{al_u}) contains arbitrary function 
$\theta(\tilde y-\tilde x)$ and real valued exponents having crucial influence on its dynamics. 
When the exponents are absent ($r=0$), this solution is nothing but the usual lump propagating along $y$ axis without a change of its form. 
If $r\ne0$, then solution (\ref{al_phi}), (\ref{al_u}) describes an appearance 
($\lambda>0$) or disappearance ($\lambda<0$) of such lump. 

The profiles of the long-wave component $u$ of the disappearing lump for different values of $\tilde\tau$ are presented in Fig.~1.
Corresponding profiles of the absolute value of the short-wave component $\varphi_1$ are given in Fig.~2. 
The dynamics of $u$ and one of the absolute value of the short-wave components are different (compare Figs.~1d, 1e with Figs.~2d, 2e). 
At the same time, the dynamics of $|\varphi_1|^2+|\varphi_2|^2$ is very similar to that of 
the long-wave component $u$. 
The structures shown in Figs.~2d, 2e will be discussed below. 

\begin{figure}[ht]
\centering
\includegraphics[width=4.0in]{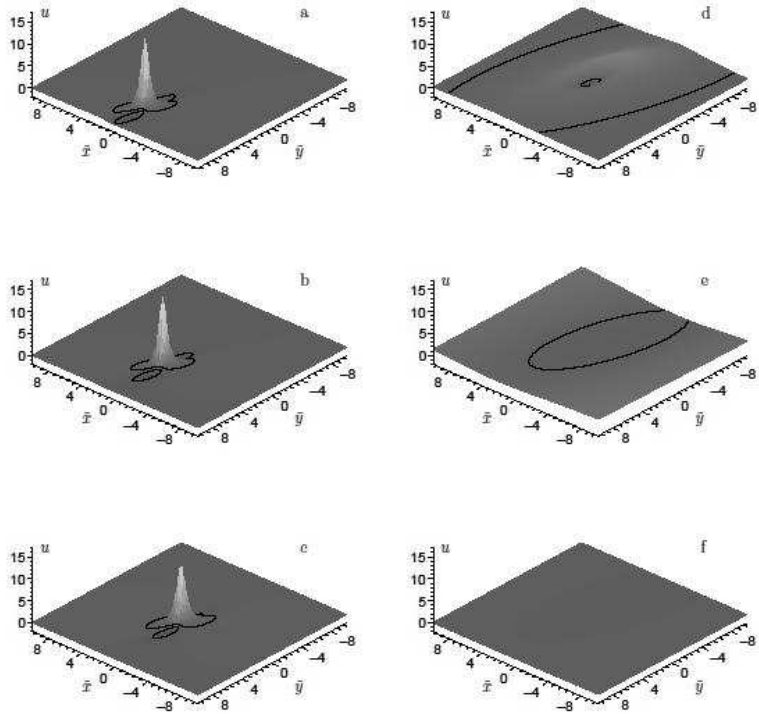}
\caption{Profiles of variable $u$ of the disappearing lump with parameters $a_1=a_2=1$, 
$r=1$, $\theta(y-x)=0$, $\lambda=-1$ and $\tilde\tau=-6$ (a), $\tilde\tau=-3.6$ (b), $\tilde\tau=-1.2$ (c), $\tilde\tau=1.2$ (d), $\tilde\tau=3.6$ (e), $\tilde\tau=6$ (f).} 
\end{figure}

\begin{figure}[ht]
\centering
\includegraphics[width=4.0in]{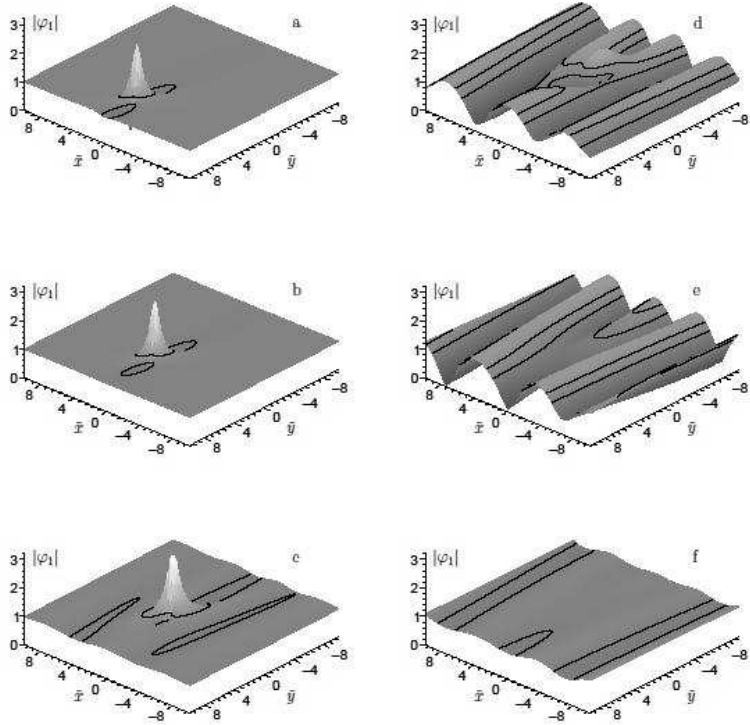}
\caption{Profiles of the absolute value of variable $\varphi_1$ of the disappearing lump with parameters as in Fig.~1 and $\tilde\tau=-6$ (a), $\tilde\tau=-3.6$ (b), 
$\tilde\tau=-1.2$ (c), $\tilde\tau=1.2$ (d), $\tilde\tau=3.6$ (e), $\tilde\tau=6$ (f).} 
\end{figure}

The process of the appearance of the lump can be understood from Figs.~1 and 2.
Note, that the phases of the short-wave components after the disappearance (appearance) of the lump differs of $\pi$ from the initial ones. 

It is seen from Eqs.~(\ref{al_phi}) that the short-wave components $\varphi_1$ and $\varphi_2$ are not proportional for any real values of parameters $a_1$ and $a_2$ if $r\ne0$.  
For this reason, solution (\ref{al_phi}), (\ref{al_u}) is not reduced to the solution of the scalar two-dimensional YO system except for the trivial case $r=0$. 

The formulas (\ref{DT_phi_L}), (\ref{DT_u_L}) allows us to construct the solutions describing an interaction of appearing and/or disappearing lumps. 
However, an existence of the lumps that appear only or disappear only rises a problem of  searching the solutions of the two-dimensional YO system (\ref{N_YO}) that join both the types of dynamics (namely, appearing-disappearing lumps). 
To solve this problem, we will use the direct approach. 

Given the formulas (\ref{al_phi}), (\ref{al_u}) and expressions for $\rho_n$ ($n=0,1,2$) and $\rho_{\pm}$, we apply the following ansatz to find the appearing-disappearing lump: 
\begin{equation}
\displaystyle\varphi_n=a_n-2\frac{Q_n+r_1R_{n}^{(-)}{\rm e}^{-\lambda\tilde\tau+i\lambda^2\tilde x}+r_2R_{n}^{(+)}{\rm e}^{\lambda\tilde\tau+i\lambda^2\tilde x}}{Q_0+|r_1|^2{\rm e}^{-2\lambda\tilde\tau}+|r_2|^2{\rm e}^{2\lambda\tilde\tau}},\ n=1,2, 
\label{adl_phi}
\end{equation}
\begin{equation}
u=2\frac{\partial^2}{\partial \tau^2}\log\left(Q_0+|r_1|^2{\rm e}^{-2\lambda\tilde\tau}+|r_2|^2{\rm e}^{2\lambda\tilde\tau}\right),
\label{adl_u}
\end{equation}
where $r_1$ and $r_2$ are the arbitrary constants, $Q_n$ ($n=0,1,2$) and $R_n^{(\pm)}$ ($n=1,2$) are the polynomials of the second and first degree of variables $\tilde\tau$, $\tilde y$ and $\tilde x$, respectively: 
\begin{equation}
\begin{array}{c}
{Q_n=b_n{\tilde\tau}^2+c_n{\tilde y}^2+d_n{\tilde x}^2+e_n\tilde\tau\tilde y+
f_n\tilde\tau\tilde x+\mbox{}}_{\mathstrut}\\
{g_n\tilde y\tilde x+h_n\tilde\tau+i_n\tilde y+j_n\tilde x+k_n,\ 
n=0,1,2,}^{\mathstrut}
\end{array}
\label{Q_n}
\end{equation}
\begin{equation}
R_n^{(\pm)}=A_n^{(\pm)}\tilde\tau+B_n^{(\pm)}\tilde y+C_n^{(\pm)}\tilde x+D_n^{(\pm)},
\ n=1,2,
\label{R_n}
\end{equation}
whose yet unknown coefficients have to be determined. 
This ansatz corresponds to the case $\theta(\tilde y-\tilde x)=0$. 

Substituting the expressions (\ref{adl_phi}) and (\ref{adl_u}) into the YO system 
(\ref{N_YO}), collecting the coefficients of monomials $\tilde\tau$, $\tilde y$, $\tilde x$ and different exponents, equalizing each coefficient in the left- and right-hand sides, we obtain the overdetermined system of the algebraic equations to define the unknowns of the polynomials $Q_n$ ($n=0,1,2$) and $R_n^{(\pm)}$ ($n=1,2$). 
It was found that this overdetermined system has the following solution: 
$$
b_0=1,\ c_0=\frac{(a_1^2+a_2^2)^2|s|^2}{4\rho\lambda^4},\ 
d_0=\frac{\sigma^2}{4\lambda^4\rho}, 
$$
$$
e_0=-f_0=\frac{\sigma(a_1^2+a_2^2)}{\lambda^2\rho}[1+2\lambda^2(r_1r_2^*+r_1^*r_2)],
$$
$$
g_0=\frac{\sigma(a_1^2+a_2^2)}{2\lambda^4\rho}[8i\lambda^5(r_1r_2^*-r_1^*r_2)-
(a_1^2+a_2^2)(1-16\lambda^4|r_1r_2|^2)],
$$
$$
h_0=i_0=j_0=0,
$$
$$
k_0=\frac{1+16\lambda^4|r_1r_2|^2}{4\lambda^2},
$$
$$
b_n=a_nb_0,\ c_n=a_nc_0,\ d_n=a_nd_0,\ e_n=a_ne_0,\ 
$$
$$
f_n=a_nf_0,\ g_n=a_ng_0,\ h_n=0,\ 
$$
$$
i_n=ia_n\frac{2d_0+g_0-8\lambda^2}{4\lambda^2},
$$
$$
j_n=-ia_n\frac{2d_0+g_0}{4\lambda^2},
$$
$$
k_n=a_n\frac{16\lambda^4|r_1r_2|^2-1}{4\lambda^2},
$$
$$
n=1,2,
$$
$$
A_1^{(\pm)}=-a_2\frac{2\lambda e_0\mp i(2d_0+g_0)}{4\lambda\sqrt{d_0}},
$$
$$
B_1^{(\pm)}=a_2\frac{g_0\pm 2i\lambda e_0}{2\sqrt{d_0}},\ C_1^{(\pm)}=a_2\sqrt{d_0},
$$
$$
D_1^{(+)}=a_2\frac{2\lambda(1-4r_1r_2^*\lambda^2)e_0-i(1+4r_1r_2^*\lambda^2)(2d_0+g_0)}
{8\lambda^2\sqrt{d_0}},
$$
$$
D_1^{(-)}=-a_2\frac{2\lambda(1-4r_1^*r_2\lambda^2)e_0+i(1+4r_1^*r_2\lambda^2)(2d_0+g_0)}
{8\lambda^2\sqrt{d_0}},
$$
$$
A_2^{(\pm)}=a_1\frac{2\lambda e_0\mp i(2d_0+g_0)}{4\lambda\sqrt{d_0}},
$$
$$
B_2^{(\pm)}=-a_1\frac{g_0\pm 2i\lambda e_0}{2\sqrt{d_0}},\ C_2^{(\pm)}=-a_1\sqrt{d_0},
$$
$$
D_2^{(+)}=-a_1\frac{2\lambda(1-4r_1r_2^*\lambda^2)e_0-i(1+4r_1r_2^*\lambda^2)(2d_0+g_0)}
{8\lambda^2\sqrt{d_0}},
$$
$$
D_2^{(-)}=a_1\frac{2\lambda(1-4r_1^*r_2\lambda^2)e_0+i(1+4r_1^*r_2\lambda^2)(2d_0+g_0)}
{8\lambda^2\sqrt{d_0}},
$$
where
$$
\sigma=16\lambda^6+(a_1^2+a_2^2)^2(1-16\lambda^4|r_1r_2|^2),
$$
$$
\begin{array}{c}
{\rho=16\lambda^6+(a_1^2+a_2^2)^2(1+16\lambda^4|r_1r_2|^2)+4(a_1^2+a_2^2)
\times\mbox{}}_{\mathstrut}\\
{\lambda^2[(a_1^2+a_2^2)(r_1r_2^*+r_1^*r_2)+4i\lambda^3(r_1r_2^*-r_1^*r_2)],}^{\mathstrut}
\end{array}
$$
$$
s=4\lambda^3(1+4\lambda^2r_1r_2^*)+i(a_1^2+a_2^2)(1-16\lambda^4|r_1r_2|^2).
$$

The substitution of the expressions written above into Eqs.~(\ref{adl_phi})--(\ref{R_n}) gives the solution of the two-dimensional YO system (\ref{N_YO}) in the form of the 
appearing-disappearing lump. 
If $r_1=0$ (or $r_2=0$), then this solution is nothing but the appearing (disappearing) lump considered in the beginning of this section. 

Note that the ansatz similar to that for the long-wave component (\ref{adl_u}), (\ref{Q_n}) was used in Refs.~\cite{LWCL}, \cite{PTZ} and \cite{WTQDZZ} in the cases of the 
(2+1)-dimensional Sawada--Kotera equation, Kadomtsev--Petviashvili equation and (2+1)-di\-mensional reduced Yu--Toda--Sasa--Fukuyama equation, respectively. 
The corresponding solutions of these equations describe an interaction of the lump with a pair of line solitons. 

The profiles of the long-wave component $u$ and the absolute value of the short-wave component $\varphi_1$ of the appearing-disappearing lump for different values of $\tilde\tau$ are presented in Fig.~3 and 4, respectively.
The duration of the interval on the variable $\tau$ axis between the end of the appearance of the lump and the beginning of its disappearance (i.e., the lump lifetime) is roughly estimated as 
\begin{equation}
\tau^*=\frac{1}{\lambda}\log\frac{k_0}{2|r_1r_2|}.
\label{lt}
\end{equation}
The amplitude of variable $u$ of the developed lump is equal approximately to 
$$
u_{max}=16\lambda^2
$$
(see Figs.~3d and 3e).

\begin{figure}[ht]
\centering
\includegraphics[width=4.0in]{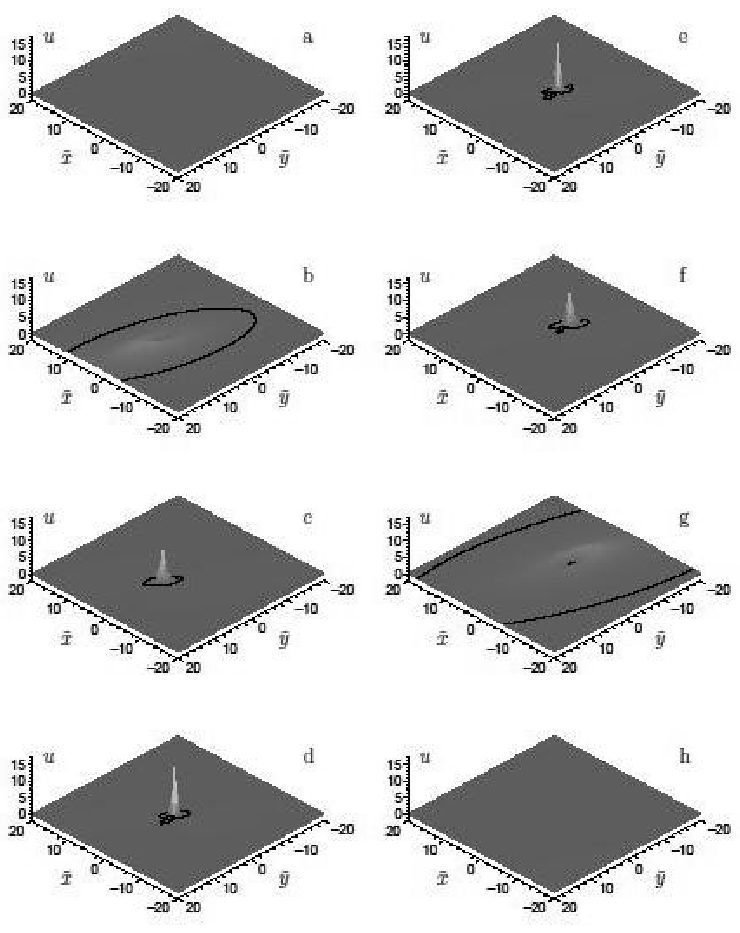}
\caption{Profiles of variable $u$ of the appearing-disappearing lump with parameters $a_1=a_2=1$, $r_1=r_2=0.01$, $\lambda=1$ and $\tilde\tau=-12$ (a), $\tilde\tau=-6$ (b), $\tilde\tau=-4$ (c), $\tilde\tau=-1$ (d), $\tilde\tau=1$ (e), $\tilde\tau=4$ (f), $\tilde\tau=6$ (g), $\tilde\tau=12$ (h).} 
\end{figure}

\begin{figure}[ht]
\centering
\includegraphics[width=4.0in]{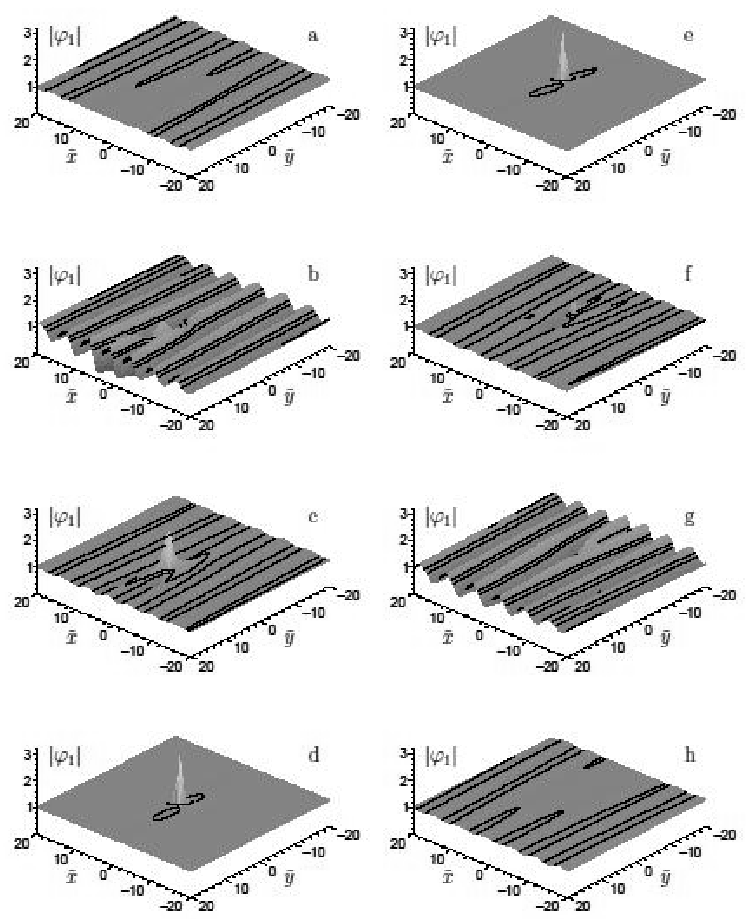}
\caption{Profiles of the absolute value of variable $\varphi_1$ of the 
appearing-disappearing lump with parameters as in Fig.~3 and $\tilde\tau=-12$ (a), $\tilde\tau=-6$ (b), $\tilde\tau=-4$ (c), $\tilde\tau=-1$ (d), $\tilde\tau=1$ (e), $\tilde\tau=4$ (f), $\tilde\tau=6$ (g), $\tilde\tau=12$ (h).} 
\end{figure}

To clarify the structures displayed in Figs.~4b, 4g (and Figs.~2d, 2e) consider the profiles given in Fig.~5 of the long-wave component $u$ and the absolute value of the short-wave component $\varphi_1$ of the appearing-disappearing lump for the fixed value of $\tilde y$.
It is seen that the lump appearance (disappearance) is accompanied by the disappearance  (appearance) of two wave packets. 
The distance between these wave packets along $\tilde x$ axis is proportional to 
$\exp(|\lambda\tilde\tau|)$. 
The minimal interval between the wave packets along $\tilde\tau$ axis is about $\tau^*$.
Note that the wave packets considered here are bright-dark (see Fig.~5b), while the solitons studied in Ref.~\cite{RPHK} are dark. 
It is remarkable that the profiles of the sum of the squares of the absolute values of the short-wave components reveal no the disappearing (or appearing) wave packets (see Fig.~5c). 

\begin{figure}[ht]
\centering
\includegraphics[width=2.0in]{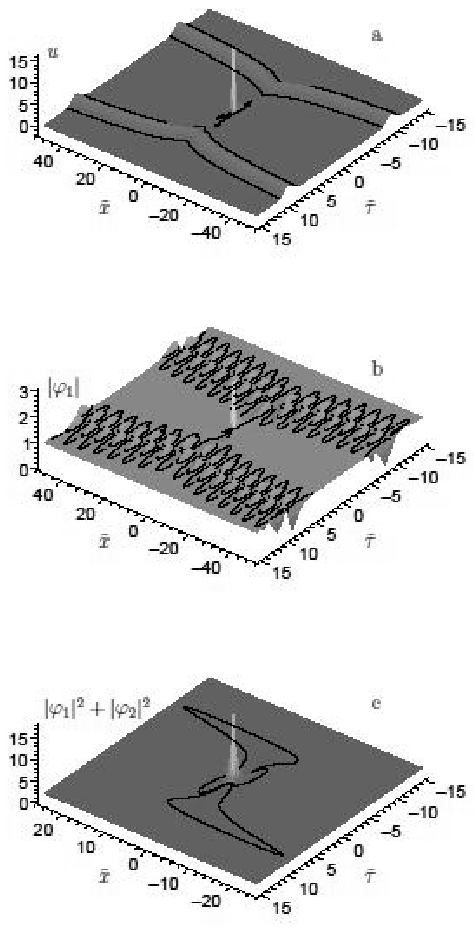}
\caption{Profiles of variable $u$ (a), the absolute value of variable $\varphi_1$ (b) and the sum of the squares of the absolute values of variables $\varphi_1$ and $\varphi_2$ (c) of the appearing-disappearing lump with parameters as in Fig.~3 and $\tilde y=0$.} 
\end{figure}

The amplitude of variable $u$ of the wave packets is equal to $u_{max}/8$, i.e., it is less on the order almost than $u_{max}$ (see Fig.~5a). 
The amplitude of $|\varphi_1|$ of the developed lump is greater than the one of the wave packets in two times approximately (Fig.~5b). 
So, the amplitudes of the appearing-disappearing lump exceed ones of the preceding and subsequent wave packets considerably. 
This feature is typical for rogue waves \cite{KPS,ORBMA,DDEG,Mih,CBSGM,CCMFK}. 
For this reason, the appearing-disappearing lump can be called as rogue lump.  

The rogue lump has free parameters $\lambda$, $r_1$, $r_2$, $\tau_0$, $y_0$ and $x_0$. 
The first three parameters determine the velocity of the lump and its lifetime. 
The remaining ones determine the location of the lump. 

It should be noted that the solution (\ref{al_phi}), (\ref{al_u}) is not reduced to the solution of the scalar (one-component) YO system. 
This points out that the ansatz (\ref{adl_phi})--(\ref{R_n}) is inapplicable in the scalar case. 

The ansatz (\ref{adl_phi})--(\ref{R_n}) used to obtain the rogue lump corresponds to the simplest case: $P(\tilde y-\tilde x)$ is a constant, $\theta(\tilde y-\tilde x)=0$. 
One may assume that the solutions of the vector YO system in the form of the localized structures exist when these constraints are not imposed. 

\section{Conclusion}

In the present report, we applied the DT technique to the two-dimensional multicomponent YO system. 
This technique gives an infinite hierarchy of the solutions of the system considered that are expressed in the terms of the solutions of the associated overdetermined linear system. 
It is important that the solutions obtained this way have the functional arbitrariness. 

The solutions of the two-dimensional multicomponent YO system on the zero and constant backgrounds were considered. 
The special attention was paid to the two-component case. 
Here, the generalizations of the lumps (namely, appearing or disappearing lumps) were obtained under the suitable choice of the arbitrary functions. 
Moreover, the further generalization of these solutions in the form of the 
appearing-disappearing lump (or rogue lump) was found by using the proper ansatz. 
The lifetime of this lump depends on its parameters and is estimated by the relation 
(\ref{lt}). 

It follows from the consideration in Sec.~4 of the generalizations of the lumps that there exists a class of the initial conditions of the vector YO system, whose evolution leads to an appearance of the localized patterns having finite lifetime. 
The amplitudes of the long- and short-wave components of these patterns can exceed significantly the ones of the initial background. 
Such kind of the dynamics resembles that of rogue waves. 

\section{Acknowledgement} 

This work was supported by the Russian Science Foundation (Project No. 17--11--01157).


\begin{thebibliography}{99}
\bibitem{OMO} Ohta Y, Maruno K, Oikawa M. Two-component analogue of two-dimensional 
long-wave–-short-wave resonance interaction equations: a derivation and solutions. J Phys A: Math Theor 2007;40:7659--72.
\bibitem{RKLG} Radha R, Kumar CS, Lakshmanan M, Gilson C R. The collision of multimode dromions and a firewall in the two-component long-wave-short-wave resonance interaction equation. J Phys A: Math Theor 2009;42:102002. 
\bibitem{KVSL} Kanna T, Vijayajayanthi M, Sakkaravarthi K, Lakshmanan M. Higher dimensional bright solitons and their collisions in a multicomponent long wave--short wave system. J  Phys A: Math Theor 2009;42:115103.
\bibitem{SK} Sakkaravarthi K, Kanna T. Dynamics of bright soliton bound states in 
(2+1)-dimensional multicomponent long wave-short wave system. Eur Phys J Spec Top 2013;222:641--653.
\bibitem{KVL} Kanna T, Vijayajayanthi M, Lakshmanan M. 2012 Mixed solitons in (2+1) dimensional multicomponent long-wave--short-wave system. Phys Rev E 2014;90:042901. 
\bibitem{KKT} Khare A, Kanna T, Tamilselvan K. Elliptic waves in two-component 
long-wave--short-wave resonance interaction system in one and two dimensions. Phys Lett A 2014;378:3093--101. 
\bibitem{CCFM_1} Chen JC, Chen Y, Feng BF, Maruno KI. Multi-dark soliton solutions of the two-dimensional multi-component Yajima--Oikawa systems. J Phys Soc Jpn 2015;84;034002. 
\bibitem{CCFM_2} Chen J, Chen Y, Feng BF, Maruno KI. Rational solutions to two- and 
one-dimensional multicomponent Yajima--Oikawa systems. Phys Lett A 2015;379:1510--9.
\bibitem{CCFMa} Chen JC, Chen Y, Feng BF, Ma ZY. General bright-dark soliton solution to 
(2 + 1)-dimensional multi-component long-wave-short-wave resonance interaction system. Nonlinear Dyn 2017;88;1273--88.
\bibitem{RPHK} Rao J, Porsezian K, He J, Kanna T. Dynamics of lumps and dark-dark solitons in the multi-component long-wave--short-wave resonance interaction system. Proc R Soc A: Math Phys Eng Sci 2018;474:20170627.
\bibitem{SU4} Sazonov SV, Ustinov NV. Two-dimensional dynamics of solitons under the conditions of Zakharov--Benney resonance. Bulletin of RAS: Physics 2018;82:1359--62.
\bibitem{ZMNP} Zakharov VE, Manakov SV, Novikov SP, Pitaevskii LP. Theory of Solitons: The
Inverse Scattering Method. New York: Consultants Bureau; 1984.
\bibitem{OOF} Oikawa M, Okamura M, Funakoshi M. Two-dimensional resonant interaction between long and short waves. J Phys Soc Jpn 1989;58:4416--30. 
\bibitem{OOS} Onorato M, Osborne AR, Serio M. Modulational instability in crossing sea states: a possible mechanism for the formation of freak waves. Phys Rev Lett 2006;96:014503.
\bibitem{SKEMS} Shukla PK, Kourakis I, Eliasson B, Marklund M, Stenflo L. Instability and evolution of nonlinearly interacting water waves. Phys Rev Lett 2006;97:094501.
\bibitem{TWKBFB} Trippenbach M, Wasilewski W, Kruk P, Bryant GW, Fibich G, Band YB. An improved nonlinear optical pulse propagation equation. Opt Comm 2002;210:385--91.
\bibitem{MWT} Matuszewski M, Wasilewski W, Trippenbach M, Band YB. Self-consistent treatment of the full vectorial nonlinear optical pulse propagation equation in an isotropic medium. Opt Comm 2003;221:337--351. 
\bibitem{CL} Cao Long V. Propagation technique for ultrashort pulses. Rev Adv Mater Sci 2010;23:8--20.
\bibitem{APT} Ablowitz MJ, Prinari B, Trubatch DA. Discrete and continuous nonlinear Schr\"odinger systems. Cambridge: Cambridge University Press; 2004.
\bibitem{KST} Kanna T, Sakkaravarthi K, Tamilselvan K. General multicomponent 
Yajima--Oikawa system: Painlev\'e analysis, soliton solutions, and energy-sharing collisions. Phys Rev E 2013;88:062921. 
\bibitem{CCKGD} Chan HN, Chow KW, Kedziora DJ, Grimshaw RHJ, Ding E. Rogue waves for a long wave-short wave resonance model with multiple short waves. Phys Rev E 2014;89:032914. 
\bibitem{CCFM} Chen JC, Chen Y, Feng BF, Maruno KI. General mixed multi-soliton solutions to one-dimensional multicomponent Yajima--Oikawa system. J Phys Soc Jpn 2015;84:074001. 
\bibitem{EPV} Estevez PG, Prada J, Villarroel J. On an algorithmic construction of lump solution in a (2+1) integrable equation. J Phys A: Math Theor 2007;40:7213--7231. 
\bibitem{KRL} Kumar CS, Radha R, Lashmanan M. Trilinearization and localized coherent structures and periodic solutions for the (2+1) dimensional KdV and NNV equations. Chaos,
Solitons \& Fractals 2009;39:942--955. 
\bibitem{DM_1} Dubard P, Matveev VB. Multi-rogue waves solutions to the focusing NLS equation and the KP-I equation. Nat Hazards Earth Syst Sci 2011;11:667--72.
\bibitem{OY_1} Ohta Y, Yang JK. Rogue waves in the Davey--Stewartson equation. Phys Rev E 2012;86:036604. 
\bibitem{OY_2} Ohta Y, Yang JK. Dynamics of rogue waves in the Davey--Stewartson II equation. J Phys A: Math Theor 2012;46:105202.
\bibitem{DM_2} Dubard P, Matveev VB. Multi-rogue waves solutions: from the NLS to the KP-I equation. Nonlinearity 2013;26:93--125.
\bibitem{WD} Wang CJ, Dai ZD. Dynamic behaviors of bright and dark rogue waves for the (2+1) dimensional Nizhnik--Novikov--Veselov equation. Phys Scr 2015;90:065205.
\bibitem{CZ} Chang L, Zeping W. Rogue waves in the (2+1) dimensional nonlinear Schr\"odinger equations. Int J Num Meth for Heat and Fluid Flow 2015;25;656--664. 
\bibitem{BWK} Baronio F, Wabnitz S, Kodama Y. Optical Kerr spatiotemporal dark-lump dynamics of hydrodynamic origin. Phys Rev Lett 2016:116;173901.
\bibitem{EDDACD} Est\'evez PG, D\'iaz E, Dom\'inguez-Adame F, Cerver\'o JM, Diez E. Lump solitons in a higher-order nonlinear equation in 2+1 dimensions. Phys Rev E  2016;93:062219. 
\bibitem{CGMB} Chen SH, Grelu P, Mihalache D, Baronio F. Families of rational soliton solutions of the Kadomtsev--Petviashvili equation. Rom Rep Phys 2016;68;1407--1424.  
\bibitem{W} Wang C. Spatiotemporal deformation of lump solution to (2+1)-dimensional KdV equation.  Nonlinear Dyn 2016;84:697--702. 
\bibitem{TD} Tan W, Dai ZD. Dynamics of kinky wave for (3+1)-dimensional potential 
Yu--Toda--Sasa--Fukuyama equation. Nonlinear Dyn  2016;85:817--23. 
\bibitem{MZD} Ma WX, Zhou Y, Dougherty R. Lump-type solutions to nonlinear differential equations derived from generalized bilinear equations. J Mod Phys B  2016;30:1640018. 
\bibitem{WZ} Wen LL, Zhang HQ. Rogue wave solutions of the (2+1)-dimensional derivative nonlinear Schr\"odinger equation. Nonlinear Dyn 2016;86;877--889. 
\bibitem{WDL} Wang CJ, Dai ZD, Liu CF. Interaction between kink solitary wave and rogue wave for (2+1)-dimensional Burgers equation. Mediterr J Math 2016;13:1087–1098. 
\bibitem{RPH} Rao JG, Porsezian K, He JS. Semi-rational solutions of the third-type 
Davey--Stewartson equation. Chaos  2017;27;083115.
\bibitem{AERS} Albares P, Estevez PG, Radha R, Saranya R. Lumps and rogue waves of generalized Nizhnik--Novikov--Veselov equation. Nonlinear Dyn 2017;90:2305--15. 
\bibitem{LWCL} Li X, Wang Y, Chen M, Li B. Lump solutions and resonance stripe solitons to the (2+1)-dimensional Sawada--Kotera equation. Adv Math Phys 2017;2017:1743789. 
\bibitem{L} Liu W. Rogue waves of the (3+1)-dimensional potential Yu--Toda--Sasa--Fukuyama equation. Rom Rep Phys 2017;69: Article no. 114. 
\bibitem{CMH} Cao Y, Malomed  BA, Hea JS. Two (2+1)-dimensional integrable nonlocal nonlinear Schr\"odinger equations: breather, rational and semi-rational solutions. 
Chaos, Solitons \& Fractals 2018;114:99--107. 
\bibitem{MS} Matveev VB, Smirnov AO. AKNS and NLS hierarchies, MRW solutions, ${\rm P}_n$ breathers, and beyond. J Math Phys 2018;59:091419. 
\bibitem{HRA} Hossen MB, Roshid HO, Ali MZ. Characteristics of the solitary waves and rogue waves with interaction phenomena in a (2+1)-dimensional Breaking Soliton equation. Phys Lett A 2018;382;1268--1274. 
\bibitem{PTZ} Peng WQ, Tian SF, Zhang TT. Analysis on lump, lumpoff and rogue waves with predictability to the (2+1)-dimensional B-type Kadomtsev--Petviashvili equation. Phys Lett A 2018;382:2701--2708. 
\bibitem{MZ} Ma WX, Zhou Y. Lump solutions to nonlinear partial differential equations via Hirota bilinear forms. J Differ Equ 2018;264:2633--59. 
\bibitem{RMCH} Rao JG, Mihalache D, Cheng Y, He JS. Lump-soliton solutions to the Fokas system. Phys Lett A 2019;383:1138--42. 
\bibitem{ZWZF} Zhou Y, Wang C, Zhang X, Fang H. Bilinear representations and lump-type waves for a fifth-order nonlinear wave equation. Eur Phys J Plus 2019;134:569. 
\bibitem{WTQDZZ} Wang M, Tian B, Qu QX, Du XX, Zhang CR, Zhang Z. Lump, lumpoff and rogue waves for a (2+1)-dimensional reduced Yu--Toda--Sasa--Fukuyama equation in a lattice or liquid. Eur Phys J Plus 2019;134:578. 
\bibitem{LWW} Liu YQ, Wen XY, Wang DS. Novel interaction phenomena of localized waves in the generalized (3+1)-dimensional KP equation. Comput Math Appl 2019;78:1--19. 
\bibitem{TDY} Tan W, Dai ZD, Yin ZY. Dynamics of multi-breathers, N-solitons and M-lump solutions in the (2+1)-dimensional KdV equation. Nonlinear Dyn 2019;96:1605--14. 
\bibitem{CW} Chen AH, Wang FF. Fissionable wave solutions, lump solutions and interactional solutions for the (2+1)-dimensional Sawada--Kotera equation. Phys Scr 2019;94:055206. 
\bibitem{KPS} Kharif C, Pelinovsky E, Slunyaev A. Rogue waves in the ocean. Berlin: Springer; 2009.
\bibitem{ORBMA} Onorato M, Residori S, Bortolozzo U, Montina A, Arecchi FT. Rogue waves and their generating mechanisms in different physical contexts. Phys Rep 2013;528:47--89.
\bibitem{DDEG} Dudley JM, Dias F, Erkintalo M, Genty G. Instabilities, breathers and rogue waves in optics. Nat Photonics 2014;8:755--64.
\bibitem{Mih} Mihalache D. Localized structures in nonlinear optical media: a selection of recent studies. Rom Rep Phys 2015;67:1383--400.
\bibitem{CBSGM} Chen SH, Baronio F, Soto-Crespo J M, Grelu P, Mihalache D. Versatile rogue waves in scalar, vector, and multidimensional nonlinear systems. J Phys A: Math Theor  2017;50:463001. 
\bibitem{CCMFK} Charalampidis EG, Cuevas-Maraver J, Frantzeskakis DJ, Kevrekidis PG. Rogue waves in ultracold bosonic seas. Rom Rep Phys 2018;70; Article no. 504. 
\bibitem{MaSa} Matveev VB, Salle MA. Darboux transformations and solitons. 
Berlin--Heidelberg: Springer--Verlag; 1991. 
\bibitem{GHZ} Gu C, Hu A, Zhou Z. Darboux Transformations in Integrable Systems. Dordrecht: Springer Science and Business Media; 2005. 
\bibitem{SU2} Sazonov SV, Ustinov NV. Propagation of vector solitons in a quasi-resonant medium with Stark deformation of quantum states. JETP 2012;115:741--58. 
\bibitem{SU3} Sazonov SV, Ustinov NV. Vector acoustic solitons from the coupling of long and short waves in a paramagnetic crystal. Theor Math Phys 2014;178:202--22. 
\bibitem{C} Chen SH. Darboux transformation and dark rogue wave states arising from two-wave resonance interaction. Phys Lett A 2014;378:1095--1098.
\bibitem{CGSC} Chen SH, Grelu P, Soto-Crespo JM. Dark- and bright-rogue-wave solutions for media with long-wave--short-wave resonance. Phys Rev E 2014;89:011201.
\bibitem{M} Melnikov VK. On equations for wave interactions. Lett Math Phys 1983;7:129-36.
\bibitem{MPK} Myrzakulov R, Pashaev OK, Kholmurodov KT. Particle-like excitations in many component magnon-phonon systems. Phys Scr 1986;33:378--84. 
\bibitem{SU1} Sazonov SV, Ustinov NV. Vector solitons generated by the long wave--short
wave interaction. JETP Lett 2011;94:610--5.
\bibitem{BS} Bugai AN, Sazonov SV. Generation of an acoustic supercontinuum under conditions of the hypersound intrapulse scattering mode. JETP 2011;112:401--13. 
\end{thebibliography}
\end{document}